\documentclass[preprint2,twocolumn,times,tighten]{aastex63}

\usepackage{color}
\usepackage{enumitem}
\usepackage{hyperref}
\usepackage{verbatim}
\usepackage{amsmath,amstext,mathrsfs}
\usepackage[all]{hypcap} 
\usepackage{afterpage}    
\usepackage{marginnote}
\usepackage{makecell}
\usepackage{subfigure}
\usepackage{multirow}
\usepackage{lineno}

\shorttitle{Anisotropies in Compressible MHD Turbulence}
\shortauthors{Hu, Xu \& Lazarian}
\graphicspath{{./}{figures/}}

\begin{document}

\title{Anisotropies in Compressible MHD Turbulence: Probing Magnetic Fields and Measuring Magnetization}

\email{yue.hu@wisc.edu; sxu@ias.edu; alazarian@facstaff.wisc.edu}

\author[0000-0002-8455-0805]{Yue Hu}
\affiliation{Department of Physics, University of Wisconsin-Madison, Madison, WI 53706, USA}
\affiliation{Department of Astronomy, University of Wisconsin-Madison, Madison, WI 53706, USA}
\author[0000-0002-5771-2055]{Siyao Xu}
\affiliation{Institute for Advanced Study, 1 Einstein Drive, Princeton, NJ 08540, USA\footnote{Hubble Fellow}}
\author{A. Lazarian}
\affiliation{Department of Astronomy, University of Wisconsin-Madison, Madison, WI 53706, USA}

\begin{abstract}
Probing magnetic fields in the interstellar medium (ISM) is notoriously challenging. Motivated by the modern theories of magnetohydrodynamic (MHD) turbulence and turbulence anisotropy, we introduce the Structure-Function Analysis (SFA) as a new approach to measure the magnetic field orientation and estimate the magnetization. We analyze the statistics of turbulent velocities in three-dimensional compressible MHD simulations through the second-order structure functions in both local and global reference frames. In the sub-Alfv\'{e}nic turbulence with the magnetic energy larger than the turbulent energy, the SFA of turbulent velocities measured in the directions perpendicular and parallel to the magnetic field can be significantly different. Their ratio has a power-law dependence on the Alfv\'{e}n Mach number $\rm M_A$, which is inversely proportional to the magnetic field strength. We demonstrate that the anisotropic structure functions of turbulent velocities can be used to estimate both the orientation and strength of magnetic fields. With turbulent velocities measured using different tracers, our approach can be generally applied to probing the magnetic fields in the multi-phase interstellar medium. 
\end{abstract}


\section{Introduction}
\label{sec:intro}
The interstellar medium (ISM) is permeated with turbulence and magnetic fields \citep{1981MNRAS.194..809L,1995ApJ...443..209A,2010ApJ...710..853C,2012ARAA...50..29,Han17,cluster}. The turbulent magnetic fields have a significant impact on diverse astrophysical processes. They are crucial in modeling the Galactic foreground \citep{2015PhRvL.115x1302C,EB,2020ApJ...899...31H}, understanding star formation \citep{2004RvMP...76..125M,MO07,2011Natur.479..499L,2012ARAA...50..29,Hu20,IRAM}, and studying cosmic rays' propagation and acceleration \citep{1949PhRv...75.1169F,1978MNRAS.182..147B,2014ApJ...783...91C,2014IJMPD..2330007B,2018ApJ...868...36X}. However, tracing the magnetic fields and characterizing the magnetized media are challenging. 

The starlight polarization and thermal emissions produced by aligned dust grains \citep{2007JQSRT.106..225L,2007ApJ...669L..77L,Aetal15} are commonly used to trace the magnetic fields in the plane-of-the-sky (POS). Using the Davis–Chandrasekhar–Fermi method \citep{1951PhRv...81..890D,1953ApJ...118..113C}, one can employ the dispersion of the directions of dust polarization and the information of spectral broadening to estimate the POS magnetic field strength in a turbulent medium \citep{2008ApJ...679..537F,2016ApJ...821...21C}. The same data can also be used to measure magnetic field with the new technique that employs differential measures instead of the dispersions (Lazarian et al. 2020). For the warm and hot phases of the ISM, the POS magnetic fields can also be measured by the synchrotron emission \citep{2006AJ....131.2900C,2016A&A...596A.103P,LP16}. Besides, the magnetic field strength along the LOS are commonly measured by molecular line splitting (Zeeman effect; \citealt{2004mim..proc..123C,2012ARAA...50..29}), which requires very high sensitivity and extremely long integration time. Faraday rotation provides another measurement for the LOS magnetic field strength \citep{1996ApJ...458..194M,2006ApJS..167..230H,2015A&A...575A.118O,2016ApJ...824..113X,2018A&A...614A.100T}. The statistics of Faraday Rotation Measure (RM), however, may be dominated by density's contribution \citep{2010ApJ...723..476A,2016ApJ...824..113X}.

Based on the development of MHD turbulence theory \citep{GS95} and turbulent reconnection theory \citep{LV99,2020PhPl...27a2305L}, several new methods have been proposed to probe the direction of magnetic fields , for instance, the Correlation Function Analysis (CFA; \citealt{2002ASPC..276..182L,2011ApJ...740..117E}), the Principal Component Analysis of Anisotropies (PCAA; \citealt{2008ApJ...680..420H}), and the more recent Velocity Gradients Technique (VGT; \citealt{2017ApJ...835...41G,YL17a,LY18a,PCA}). Some of the above methods employ the anisotropy of MHD turbulence, i.e., the turbulent eddies are elongated along the magnetic field and smaller eddied are relatively more elongated. Consequently, the eddy's orientation reveals the local magnetic field direction. Also, the magnetic field strength can be observationally estimated
based on the theoretical understanding of properties of MHD turbulence
(e.g. \citealt{2008ApJ...677.1151L,2015ApJ...814...77E,2016ApJ...833..215X,2018ApJ...865...46L,2020arXiv200207996L,2020ApJ...898...66Y}). 
In addition to the eddy's anisotropic shape, the amplitude of velocity fluctuations also shows the anisotropy \citep{2000ApJ...539..273C,2001ApJ...554.1175M,2002ApJ...564..291C,2003MNRAS.345..325C}. At the same distance away from the eddy's center, velocity fluctuations are maximum in the direction perpendicular to the local magnetic fields, as the turbulent motions, due to fast turbulent reconnection, are not constrained to mix up magnetized media in the direction perpendicular to the {\it local} direction\footnote{The notion of the local direction of magnetic field is the key concept of the MHD turbulence theory. This concept was missing in the original \citep{GS95} study. However, it trivially follows from the theory of turbulent reconnection. The Goldreich-Sridhar relation between the velocities of parallel and perpendicular to magnetic field are only valid in the system of reference related to the {\it local} direction of magnetic field. This was first demonstrated by \citep{2000ApJ...539..273C} and confirmed by the subsequent numerical studies.} of magnetic field \citep{LV99}. Such motions provide the minimal bending of magnetic field and therefore the cascade evolves in the direction of the minimum resistance. The minimum amplitude, therefore, indicates the magnetic field direction. This property provides the theoretical foundation of our proposed new approach, i.e., the Structure-Function Analysis (SFA) of turbulent velocities, in tracing the magnetic fields. 

The SFA proposed in this work employs the second-order structure function of turbulent velocities. It measures the amplitude of velocity fluctuations 
along different directions 
in the global reference frame. The direction corresponding to the minimum amplitude reveals the direction of the mean magnetic field. This idea was earlier implemented in statistical studies of velocity centroids to reveal the direction of the POS mean magnetic field \citep{2002ASPC..276..182L,2005ApJ...631..320E,2011ApJ...740..117E}. \citet{2014ApJ...790..130B} later performed a parameter study showing the degree of anisotropy, i.e., the ratio of velocity centorid fluctuations measured in the perpendicular and parallel directions (with respect to the POS mean magnetic field), is related primarily to magnetic field strength. The relation of the anisotropy degree and the magnetic field strength is analytically quantified by \citet{2017MNRAS.464.3617K}. In this work, we use 3D velocity of point sources rather than velocity centroid to calculate the degree of anisotropy. We analytically and numerically show that the degree of anisotropy has a power-law dependence on the Alfv\'{e}n Mach number $\rm M_A$ in sub-Alfv\'{e}nic turbulence. Instead of using 2D velocity centroids as an indirect measurement of velocity fluctuations, the SFA can utilize 3D velocities of point sources to trace 3D magnetic field and measure the magnetization $\rm M_A^{-1}$. 

The paper is organized as follows.  In \S~\ref{sec:theory}, we 
analyze the anisotropy of the velocity structure functions and 
discuss its differences in the local and global reference frames. 
In \S~\ref{sec:data}, we provide the details of the numerical data used in this work. 
In \S~\ref{sec:results}, we perform numerical experiments to test our analytical results. 
In \S~\ref{sec:obs}, we discuss the observational applications of SFA. 
We present discussion in \S~\ref{sec.dis} and summary in \S~\ref{sec:con}.


\section{Theoretical formulation of the SFA}
\label{sec:theory}
\subsection{Anisotropy of MHD turbulence}
\begin{figure*}
\centering
\includegraphics[width=1.0\linewidth,height=0.3\linewidth]{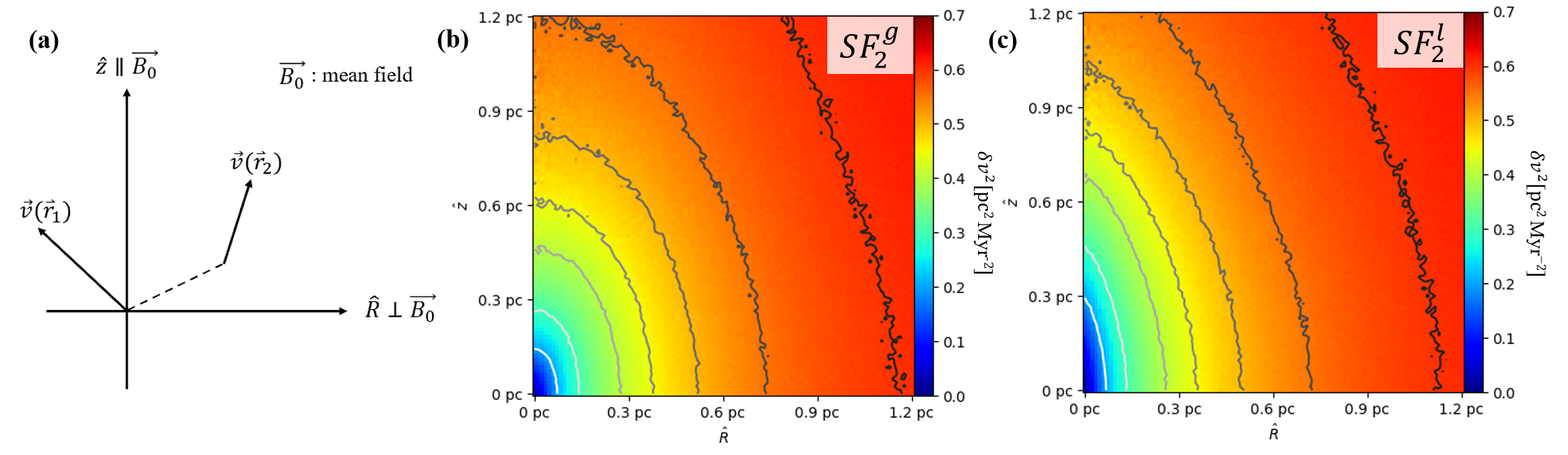}
\caption{\label{fig:illustration} \textbf{Panel (a)}: the definition of structure-function in the global reference frame. We adopt a cylindrical coordinate system in which the z-axis is parallel to the mean magnetic field $\boldsymbol{B_0}$. \textbf{Panel (b)}: an example of structure-function $SF_2^g$ in the global reference frame using simulation A7. \textbf{Panel (c)}: an example of structure-function $SF_2^l$ in the local reference frame using simulation A7. }
\end{figure*}

Several attempts to describe statistics of MHD turbulence have been achieved in decades. \citet{1963AZh....40..742I} and \citet{1965PhFl....8.1385K} proposed an isotropic model of turbulence despite the presence of the magnetic field. However, a number of theoretical and numerical studies later revealed that sub-Alfv\'{e}nic MHD turbulence is indeed anisotropic instead of isotropic \citep{1981PhFl...24..825M,1983JPlPh..29..525S,1984ApJ...285..109H,1995ApJ...447..706M}. A cornerstone of the anisotropic MHD turbulence theory was given by \citet{GS95}, denoted as GS95. By considering the "critical balance" condition, i.e., equating the cascading time ($k_\bot v_l$)$^{-1}$ and the wave periods ($k_\parallel v_A$)$^{-1}$, GS95 obtained the anisotropy scaling in the global reference frame:
\begin{equation}
    k_\parallel\propto k_\bot^{2/3}
\end{equation}
where $k_\parallel$ and $k_\bot$ are the components of the wavevector parallel and perpendicular to the magnetic field, respectively. $v_l$ is turbulent velocity at scale $l$ and $v_A$ is Alfv\'{e}n speed. The GS95 scaling indicates that turbulent eddies are elongating along the large-scale mean magnetic fields. This scale-dependent anisotropy, however, is not observable in the global reference frame. Only the largest eddy appears in the global frame due to the averaging effect \citep{2000ApJ...539..273C}. The study of fast turbulent reconnection done by \citet{LV99} demonstrated that the eddy-like description of sub-Alfv\'{e}nic turbulence is only valid if the eddies axes are aligned with the local magnetic field direction. They derived the anisotropy relation for the eddies in the local reference frame:
\begin{equation}
\label{eq.lv99}
 l_\parallel= L_{inj}(\frac{l_\bot}{L_{inj}})^{\frac{2}{3}}{\rm M_A^{-4/3}}, {\rm M_A\le 1}
\end{equation}
where $l_\bot$ and $l_\|$ are the perpendicular and parallel scales of eddies with respect to the {\it local} magnetic field, respectively. $L_{inj}$ is the turbulence injection scale. Using the "critical balance" condition expressed in the local frame, \citet{LV99} obtained the corresponding scaling of velocity fluctuation: 
\begin{equation}
\label{eq.v}
 v_l= v_{inj}(\frac{l_\bot}{L_{inj}})^{\frac{1}{3}}{\rm M_A^{1/3}}
\end{equation}
where $v_{inj}$ is the injection velocity. \citet{2000ApJ...539..273C} and \citet{2001ApJ...554.1175M} later numerically demonstrated this scale-dependent anisotropy in the local reference frame.

The above picture of eddy-like sub-Alfv\'{e}nic turbulence is valid only at scales from  dissipation scale $l_{diss}$ to transition scale $l_{tr}=L_{inj}{\rm M_A^2}$ \citep{2006ApJ...645L..25L}. The turbulence within this regime [$l_{diss}$, $l_{tr}$] is called strong MHD turbulence. The turbulence in the range from $L_{inj}$ to $l_{tr}$ is termed weak Alfvénic turbulence, which is described as wave-like rather than eddy-like. The turbulent velocity at the transition scale $v_{tr}=v_{inj}{\rm M_A}$ leads to the expressions:
\begin{equation}
\label{eq.4}
\begin{aligned}
   v_l&= v_{tr}(\frac{l_\bot}{l_{tr}})^{\frac{1}{3}}\\
   l_\parallel&= {\rm M_A^{-2}}l_\bot^{\frac{2}{3}}l_{tr}^{\frac{1}{3}}
\end{aligned}
\end{equation}
which at $l_{tr}$ becomes:
\begin{equation}
    l_{tr,\parallel}=\frac{v_A}{v_{tr}}l_{tr,\bot}
\end{equation}
because of $l_{tr}\approx l_{tr,\bot}$. It shows that the anisotropy of strong MHD turbulence is scale-dependent. The velocity fluctuation $v_l$ can also be expressed as:
\begin{equation}
       v_l= v_{tr}{\rm M_A}(\frac{l_\parallel}{l_{tr}})^{\frac{1}{2}}
\end{equation}
When $l_\bot$ of one eddy (eddy 1) is equal to $l_\parallel$ of another eddy (eddy 2), the corresponding velocity fluctuations $v_{l,1}$ and $v_{l,2}$ are:
\begin{equation}
\begin{aligned}
   (\frac{l_\bot}{l_{tr}})^{\frac{2}{3}}&=v_{l,1}^2 v_{tr}^{-2}\\
   (\frac{l_\parallel}{l_{tr}})^{\frac{2}{3}}&=v_{l,2}^{4/3} v_{tr}^{-4/3}{\rm M_A}^{-4/3}\\
\end{aligned}
\end{equation}
By equating these two expressions, one gets:
\begin{equation}
\begin{aligned}
\label{eq:vr_loc}
   v_{l,1}^2/v_{l,2}^2 &= (v_{l,2}/v_{tr})^{-2/3}{\rm M_A}^{-4/3}\\
   &=(l_\parallel/l_{tr})^{-1/3}{\rm M_A}^{-2},\\
   &=(l_\parallel/L_{inj})^{-1/3}{\rm M_A}^{-4/3}, {\rm M_A\le 1}\\
\end{aligned}
\end{equation}
which is valid in the local reference frame. This ratio is called the degree of anisotropy. In the global reference frame, the observed anisotropy is determined by the anisotropy of the largest eddy at $l_{tr}$. 

In the global reference frame, we have:
\begin{equation}
    \frac{l_\parallel}{l_\bot}=\frac{l_{tr,\parallel}}{l_{tr,\bot}}=\frac{v_A}{v_{tr}}
\end{equation}
$v_l$ can be expressed in terms of $l_\parallel$:
\begin{equation}
    v_l=v_{tr}(\frac{l_\parallel}{l_{tr,\parallel}})^{\frac{1}{3}}
\end{equation}
Comparing the above expression with Eq.~\ref{eq.4}, we find that
to have $l_\parallel$ = $l_\bot$, the turbulent velocities should satisfy:
\begin{equation}
\label{eq:vr_glo}
\begin{aligned}
     v_{l,1}^2/v_{l,2}^2 &=(\frac{l_\parallel}{l_{tr,\parallel}})^{\frac{2}{3}}=(\frac{v_A}{v_{tr}})^{\frac{2}{3}}\\
     &\approx {\rm M_A}^{-4/3}, {\rm M_A\le1}
\end{aligned}
\end{equation}
It shows that the ratio measured in the global frame of mean magnetic field is a power-law relation with $\rm M_A$ for sub-Alfv\'{e}nic turbulence. Velocity fluctuation in the direction perpendicular to the magnetic field has more significant amplitude. Therefore, once the ratio $v_{l,1}^2/v_{l,2}^2$ in the global is measured, one can could get the magnetization level  ${\rm M_A^{-1}}$ from:
\begin{equation}
    {\rm M_A}^{-1}=(v_{l,1}^2/v_{l,2}^2)^{3/4}, {\rm M_A\le1}\\
\end{equation}

In super-Alfv\'{e}nic ($\rm M_A>1$) condition, the transition to strong MHD turbulence happens at scale $l_a=L_{inj}{\rm M_A^{-3}}$ with transition velocity $v_{st}=v_A$ \citep{2006ApJ...645L..25L}. Similarly, we can express $l_\parallel$ and $v_l$ in terms of $l_a$:
\begin{equation}
\begin{aligned}
l_\parallel&= {\rm M_A^{-1/3}}l_\bot^{\frac{2}{3}}l_{a}^{1/3}\\
   v_l&= v_{tr}(\frac{l_\bot}{l_{a}})^{\frac{1}{3}}{\rm M_A^{1/3}}=v_{tr}(\frac{l_\parallel}{l_{a}})^{\frac{1}{2}}{\rm M_A^{1/2}}
\end{aligned}
\end{equation}
Consequently, we can obtain the ratio $v_{l,\bot}^2/v_{l,\parallel}^2$ in the local reference frame:
\begin{equation}
   v_{l,1}^2/v_{l,2}^2 =(l_\parallel/l_{a})^{-1/3}{\rm M_A}^{-1/3}, {\rm M_A>1}
\end{equation}
which can return to Eq.~\ref{eq:vr_loc} by replacing $l_a$ with $l_{a}=L_{inj}{\rm M_A^{-3}}$. In the global reference frame, super-Alfv\'{e}nic is isotropic:
\begin{equation}
\label{eq.12}
   v_{l,1}^2/v_{l,2}^2 = 1, {\rm M_A>1}
\end{equation}
Above we follow the consideration of incompressible turbulence used by GS95 and LV99. \citet{2003MNRAS.345..325C} later confirmed that these scaling relations are still valid for compressible turbulence when the Alfv\'{e}nic mode dominates the turbulence.

\subsection{Structure functions of turbulent velocities in the local and global reference frames}
As discussed above, the ratio between the two components  $v_{l,1}^2$ and $v_{l,2}^2$ can be used to estimate the Alfv\'{e}n Mach number. To calculate $v_{l,1}^2$ and $v_{l,2}^2$, we employ the second order structure-function in both the local and global reference frames. The definition of the structure-function in the local reference frame is given by \citet{2000ApJ...539..273C}:
\begin{equation}
\label{eq.st_loc}
\begin{aligned}
    &\boldsymbol{B_l}=\frac{1}{2}(\boldsymbol{B}(\boldsymbol{r_1})-\boldsymbol{B}(\boldsymbol{r_2}))\\
    &SF_2^l(R,z)=\langle|\boldsymbol{v}(\boldsymbol{r_1})-\boldsymbol{v}(\boldsymbol{r_2})|^2 \rangle
\end{aligned}
\end{equation}
where $\boldsymbol{B_l}$ defines the local magnetic fields. R and z are coordinates in a cylindrical coordinate system in which the z-axis is parallel to $\boldsymbol{B_l}$. Explicitly, $R=|\hat{z}\times(\boldsymbol{r_1}-\boldsymbol{r_2})|$, $z=\hat{z}\cdot(\boldsymbol{r_1}-\boldsymbol{r_2})$, and $\hat{z}=\boldsymbol{B_l}/|\boldsymbol{B_l}|$. Similarly, we can replace the local magnetic fields $\boldsymbol{B_l}$ by the mean magnetic fields $\boldsymbol{B_0}$, we obtain the structure-function in the global reference frame: 
\begin{equation}
        SF_2^g(R,z)=\langle|\boldsymbol{v}(\boldsymbol{r_1})-\boldsymbol{v}(\boldsymbol{r_2})|^2 \rangle
\end{equation}
where $z=\hat{z}\cdot(\boldsymbol{r_1}-\boldsymbol{r_2})$ and $\hat{z}=\boldsymbol{B_0}/|\boldsymbol{B_0}|$, as shown in Fig.~\ref{fig:illustration}. The local structure-function is more anisotropic toward smaller scale. In both frames, $v_{l,1}^2$ and $v_{l,2}^2$ are obtained from \citep{2002ApJ...564..291C}:
\begin{equation}
\label{eq.18}
\begin{aligned}
   v_{l,1}^2&=SF_2(R,0)\\
   v_{l,2}^2&=SF_2(0,z)
\end{aligned}
\end{equation}

The scale-dependent anisotropy scaling of Eq.~\ref{eq.v} can be observed in a local reference frame, which is defined with respect to the local mean magnetic field \citep{2000ApJ...539..273C,2001ApJ...554.1175M}. In the global reference frame, the anisotropy is dictated by the largest eddy (i.e., the eddy at scale $l_{st}$ for sub-Alfv\'{e}nic turbulence) so that the observed anisotropy is scale-independent. However, due to the isotropic driving, it is possible to observe ( an artificial) scale-dependent anisotropy in the global frame of reference, which is aligned with the mean magnetic field (see \citealt{2003ApJ...590..858V,2018ApJ...865...54Y}). This artificial anisotropy comes from the isotropic driving instead of turbulence's cascade. The isotropic driving produces large-scale isotropic structures. Small-scale structures, however, are anisotropic. The structure-function in the global reference frame continuously measures the velocity fluctuations from small scales to large scales. Therefore, there should be transition scales on which we observe an artificial scale-dependent anisotropy. With sufficiently long inertial range (longer than the transitional range), we expect to see scale-independent global anisotropy on scales far smaller than the driving scale \citep{2016A&A...595A..37K}.

\section{Numerical Data}
\label{sec:data}
\begin{table}
\centering
\label{tab:sim}
\begin{tabular}{| c | c | c | c | c | c|}
\hline
Model & $\rm M_S$ & $\rm M_A$  & Resolution & $\beta$ & $l_{st}$ or $l_a$ [pc] \\\hline \hline
A1 & 0.63 & 0.37 & $792^3$ & 0.69 & 0.68\\
A2 & 0.60 & 0.78 & $792^3$ & 3.38 & 3.05\\
A3 & 0.60 & 1.02 & $792^3$ & 5.78 & 4.71\\
A4 & 1.27 & 0.50 & $792^3$ & 0.31 & 1.25\\
A5 & 5.55 & 1.71 & $792^3$ & 0.19 & 1.00\\
A6 & 10.81 & 0.26 & $792^3$ & 0.001 & 0.34\\
A7 & 10.61 & 0.68 & $792^3$ & 0.007 & 2.30\\
\hline
\end{tabular}
\caption{Description of our MHD simulations. $\rm M_S$ and $\rm M_A$ are the instantaneous values at each the snapshots are taken. The compressibility of turbulence is characterized by $\beta=2(\frac{\rm M_A}{\rm M_S})^2$.}
\end{table}
We perform 3D MHD simulations through ZEUS-MP/3D code \citep{2006ApJS..165..188H}, which solves the ideal MHD equations in a periodic box:
\begin{equation}
    \begin{aligned}
      &\partial\rho/\partial t +\nabla\cdot(\rho\Vec{v})=0\\
      &\partial(\rho\Vec{v})/\partial t+\nabla\cdot[\rho\Vec{v}\Vec{v}+(p+\frac{B^2}{8\pi})\Vec{I}-\frac{\Vec{B}\Vec{B}}{4\pi}]=f\\
      &\partial\Vec{B}/\partial t-\nabla\times(\Vec{v}\times\Vec{B})=0
    \end{aligned}
\end{equation}
where $f$ is a random large-scale driving force, $\rho$ is the density, $\Vec{v}$ is the velocity, and $\Vec{B}$ is the magnetic field. We also consider a zero-divergence condition $\nabla \cdot\Vec{B}=0$, and an isothermal equation of state $p = c_s^2\rho$, where $p$ is the gas pressure. We consider single fluid and operator-split MHD conditions in the Eulerian frame. The simulation is staggered grided to 792$^3$ cells. The turbulence is solenoidally injected at a spatial scale k $\approx$ 2. 

The MHD simulation of astrophysical turbulence is scale-free and we may assign any desired physical scales. Here we specify temperature T = 10.0 K, sound speed $c_s$ = 187 m/s and cloud size L = 10 pc to emulate an isothermal cloud. The magnetic field is considered as $\Vec{B}=\Vec{B_0}+\delta\Vec{B}$, where $\Vec{B_0}$ is the uniform background field and $\delta\Vec{B}$ is the magnetic fluctuation. $\Vec{B_0}$ is assumed to be perpendicular to the LOS. We vary the sonic Mach number M$_{\rm S}=v_{inj}/c_{s}$ and Alfv\'{e}nic Mach number M$_{\rm A}=v_{inj}/v_{A}$ to explore different physical conditions. Here $v_{inj}$ is the isotropic injection velocity and $v_{A}$ is the Alfv\'{e}nic velocity. We refer to the simulations in Tab.~\ref{tab:sim} by their model name or parameters.

\begin{figure}
\centering
\includegraphics[width=1.0\linewidth,height=1.45\linewidth]{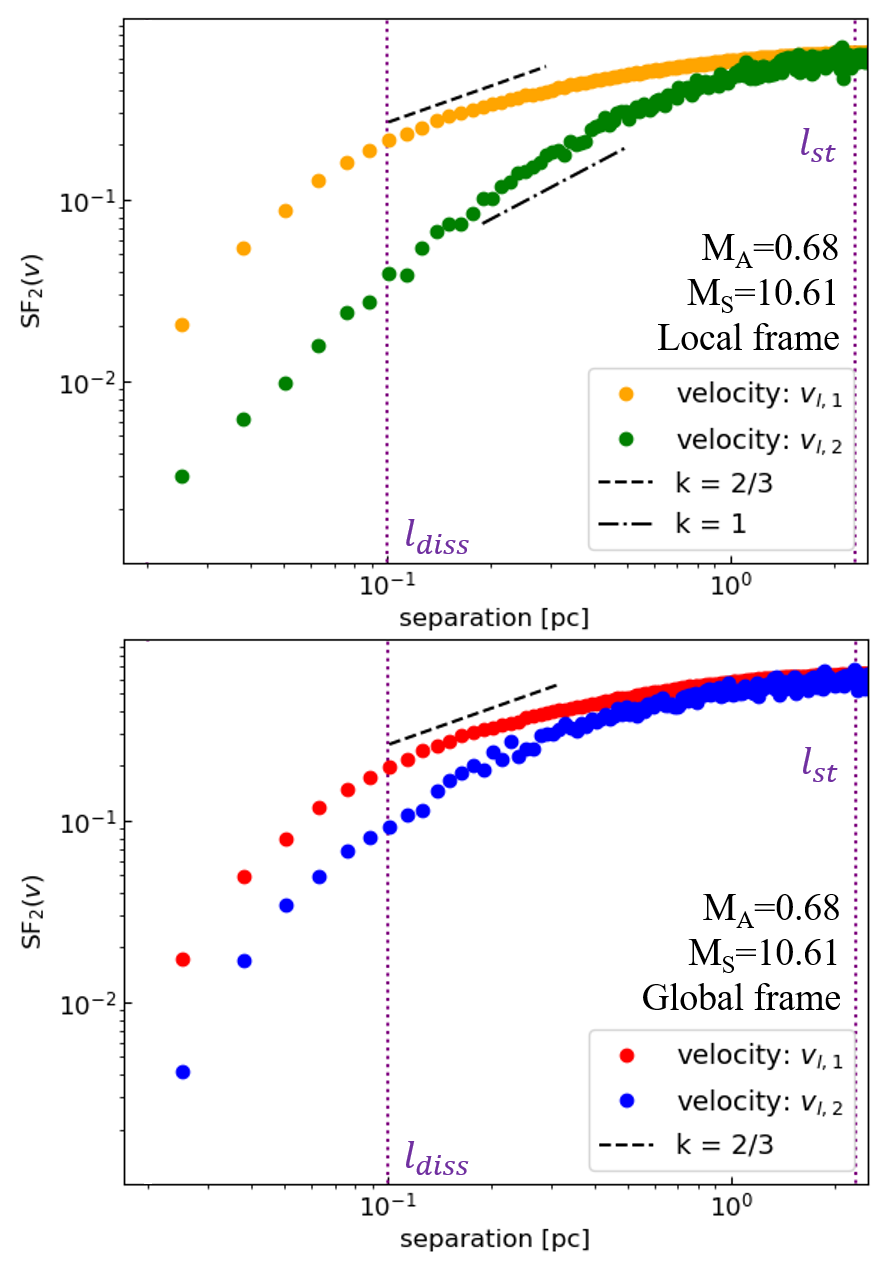}
\caption{\label{fig:SF_sub02} \textbf{Panel (a)}: the structure-function in the local reference frame. The simulation A7 is used here. \textbf{Panel (b)}: an example of structure-function in the global reference frame. k denotes the slope of the reference line. $l_{diss}$ and $l_{st}$ represent dissipation scale and transition scale, respectively.}
\end{figure}
\section{Results}
\label{sec:results}
\subsection{Probing the direction of magnetic fields}
\label{subsec:sfa}
Fig.~\ref{fig:SF_sub02} presents an example of the local structure-function and the global structure-function. We use the sub-Alfv\'{e}nic simulation A7 ($\rm M_A=0.68$ and $\rm M_S=10.61$). The corresponding sonic scale $l_s=L_{inj}{\rm M_S}^{-3}\approx4\times10^{-3}$ pc, considering Kolmogorov-type cascading. Based on the probability distribution functions of velocity, we adopt the Monte Carlo method to select one million points to calculate structure-function. In the local frame, we observe that at scales $l_{diss}\le l\le l_{st}$, the velocity field follows $v_1^2\propto l_\bot^{2/3}$ and $v_2^2\propto l_\parallel$. At scales larger than $l_{st}$, the turbulence becomes weak, i.e., wave-like, and the driving effect gets significant. This agrees with the numerical results of \citet{2000ApJ...539..273C}. In molecular cloud, the scales from $L_{inj}\approx5.0$ pc to $l_{diss}\approx0.1$ pc typically correspond to volume density of non-self-gravitating gas from $\approx 1\times10^2$ g cm$^{-3}$ to $\approx 1\times10^5$ g cm$^{-3}$ \citep{2013A&A...549A.132P}.

In real scenario, measurement of the local structure-function requires the knowledge of 3D position, 3D velocity, and 3D magnetic field distribution. This requirement is only accessible for in-situ measurements in the solar wind, where local anisotropy is indeed confirmed \citep{2016ApJ...816...15W}. The global frame, however, can be more easily accessed in observations as we will discuss. In the global frame, we also find  $v_1^2\propto l_\bot^{2/3}$, and the two velocities get overlapped when scales larger than $l_{st}$. Ideally, the structure-functions of $v^2_1$ and $v^2_2$ should exhibit identical slope since the global anisotropy is scale-independent. However, as discussed above, the isotropic driving produces isotropic velocity fluctuations so that $v^2_1$ and $v^2_2$ are transited to the same amplitude on a large scale. This transition induces an artificial anisotropy in the global reference frame. 
\begin{figure}
\centering
\includegraphics[width=1.0\linewidth,height=0.74\linewidth]{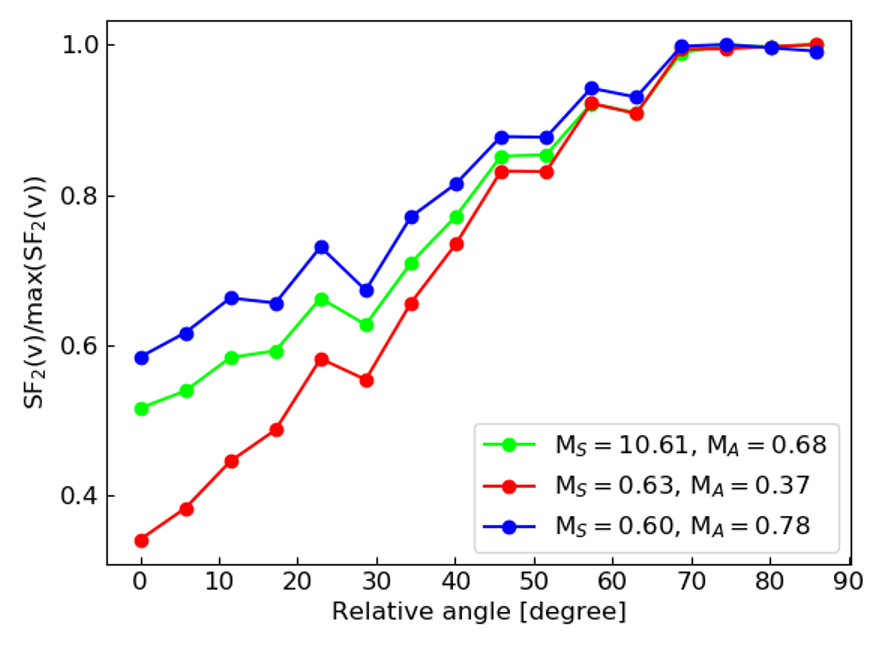}
\caption{\label{fig:angle} The correlation of normalized structure function and the relative angle between $\hat{r}$ and mean magnetic field  $\Vec{B_0}$. $SF_2(v)$ takes the value at 0.1 pc.}
\end{figure}

Additionally, $v^2_1$ consistently exhibits a higher amplitude than $v^2_2$ in both frames, which implies the turbulence cascade is preferentially along the direction perpendicular to the magnetic fields. Theoretically, it was explained by the turbulence reconnection theory \citep{LV99}. In the sub-Alfv\'enic regime, Alfv\'en waves initially evolve by the so-called weak turbulence cascade increasing the perpendicular wavenumber while keeping the parallel wavenumber the same \citep{LV99,2000JPlPh..63..447G,2005PhPl...12i2310G}. The perpendicular wavenumber's increase makes the Alfv\'enic wave vectors more and more perpendicular to the magnetic field. The weak cascade proceeds until the critical balance condition, namely, $l_\|/v_A\sim l_\bot/v_\bot$ is fulfilled. The critical balance is the cornerstone of the theory of strong turbulence \citep{GS95}. From the point of view of turbulent reconnection, the critical balance corresponds to the condition that the turbulent eddies aligned with the local direction magnetic field perform their turnover in the time $l_\bot/v_\bot$ and this time is equal to the period of the wave that the eddy motion excites, i.e., to $l_\|/v_A$. Such motions occur as the fast turbulent reconnection changes the topology of the interacting magnetic flux tubes within one eddy turnover time \citep{LV99}. Consequently, in the local reference frame, the fluid motions perpendicular to magnetic fields are not constrained by magnetic tension. This perpendicular direction, therefore, provides the path of least resistance for turbulent cascade.

Initially, we consider the situation that $\hat{r}$ direction\footnote{In Eq.~\eqref{eq.18}, we define $\hat{z}$ is parallel to $\Vec{B_0}$ and $\hat{R}$ is parallel to $\Vec{B_0}$. Here we use $\hat{r}$ to represent an arbitrary direction for calculating the structure-function.} of the velocity structure-function is parallel to the mean magnetic field $\Vec{B_0}$. Next, we vary the relative angle between $\hat{r}$ and $\Vec{B_0}$ when calculating the structure-function. In Fig.~\ref{fig:angle}, we plot the relation of measured structure-function $SF_2(v)$ and relative angle between $\hat{r}$ and $\Vec{B_0}$. To avoid the artificial anisotropy (see \S~\ref{sec:theory}), the $SF_2(v)$ takes the value at 0.1 pc, below which the turbulence starts numerically dissipating. In the real observation, one can take the $SF_2(v)$ at larger scale as there exists scale-independent global anisotropy in an extended inertial range. Consequently, one can average the anisotropy over the inertial range to reduce the uncertainty introduced by insufficient sampling. For visualization purposes, the resulting $SF_2(v)$ is normalized so that the maximum value is 1. We find generally the normalized $SF_2(v)$ is increasing when the relative angle gets large. The increment is more rapid when the relative angle is less than 70 degrees. Nevertheless, the normalized $SF_2(v)$ achieves its maximum when $\hat{r}$ is perpendicular to $\Vec{B_0}$ (i.e., $\hat{r}=\hat{R}$) and has its minimum value when $\hat{r}$ is parallel to $\Vec{B_0}$ (i.e., $\hat{r}=\hat{z}$). The direction $\hat{r}$ corresponding to minimum value of $SF_2(v)$, therefore, gives the magnetic field direction.

\subsection{Measuring magnetization}
\begin{figure}
\centering
\includegraphics[width=1.0\linewidth,height=0.77\linewidth]{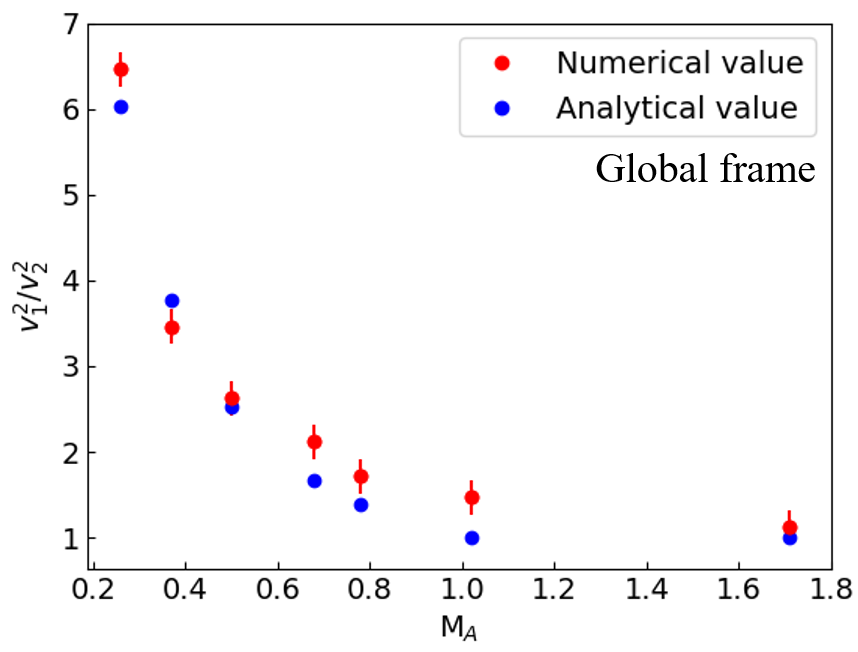}
\caption{\label{fig:Ma} The correlation of $\rm M_A$ and $v_{l,1}^2/v_{l,2}^2$. The calculation is performed in the global reference frame, selecting the velocity fluctuations at the scale $l\approx0.1$ pc, below which the turbulence starts numerically dissipating. The analytical expressions are $v_{l,1}^2/v_{l,2}^2={\rm M_A}^{-4/3}$ for $\rm M_A\le1$ and $v_{l,1}^2/v_{l,2}^2=1$ for $\rm M_A>1$. }
\end{figure}

In the above section, we discussed how to probe the magnetic field direction using the structure-function. The maximum velocity fluctuation appears in the direction perpendicular to the magnetic fields. Once the magnetic field direction is known, the corresponding ratio $v_{l,1}^2/v_{l,2}^2$ can reveal the magnetization $\rm M_A^{-1}$ (see Eq.~\ref{eq:vr_glo} and Eq.~\ref{eq.12}).

Fig.~\ref{fig:Ma} presents the correlation of $\rm M_A$ and $v_{l,1}^2/v_{l,2}^2$ in the global reference frame. To avoid this driving effect, we only consider the velocity fluctuations at the scale $l\approx0.1$ pc, below which the turbulence starts numerically dissipating. Uncertainty is given by the standard deviation of the mean. We see that the measured ratio well follows the theoretical correlation:
\begin{equation}
\label{eq.20}
   v_{l,1}^2/v_{l,2}^2=
\begin{cases}
        {\rm M_A}^{-4/3}, &{\rm (global, M_A\le1)}\\
        1, &{\rm (global, M_A>1)}\\
\end{cases}
\end{equation}
The change in power-law index for $\rm M_A>1$ cases is expected. When the injection velocity becomes higher than the Alfv\'{e}n speed, the large-scale motions are dominated by hydro-type turbulence, and the directions of the magnetic field within the flow are significantly randomized so that the power-law gets shallower. Note that in observations, the structure-function can only be calculated in the global reference frame. The measure velocity fluctuations are only correlated to $\rm M_A$ in this frame. Therefore, $\rm M_A$ can be inferred from $v_{l,1}^2/v_{l,2}^2$, which are measured by the global structure-function. We call this approach as Structure-Function Analysis (SFA). In practice, one should  measure the velocity structure-function from different position angle to find the maximum and minimum values of velocity fluctuations. The magnetic field strength can be calculated from $B_0=\sqrt{4\pi\rho}v_{inj}/{\rm M_A}$, where $\rho$ is the gas mass density.
\begin{figure*}
\centering
\includegraphics[width=1.0\linewidth,height=0.35\linewidth]{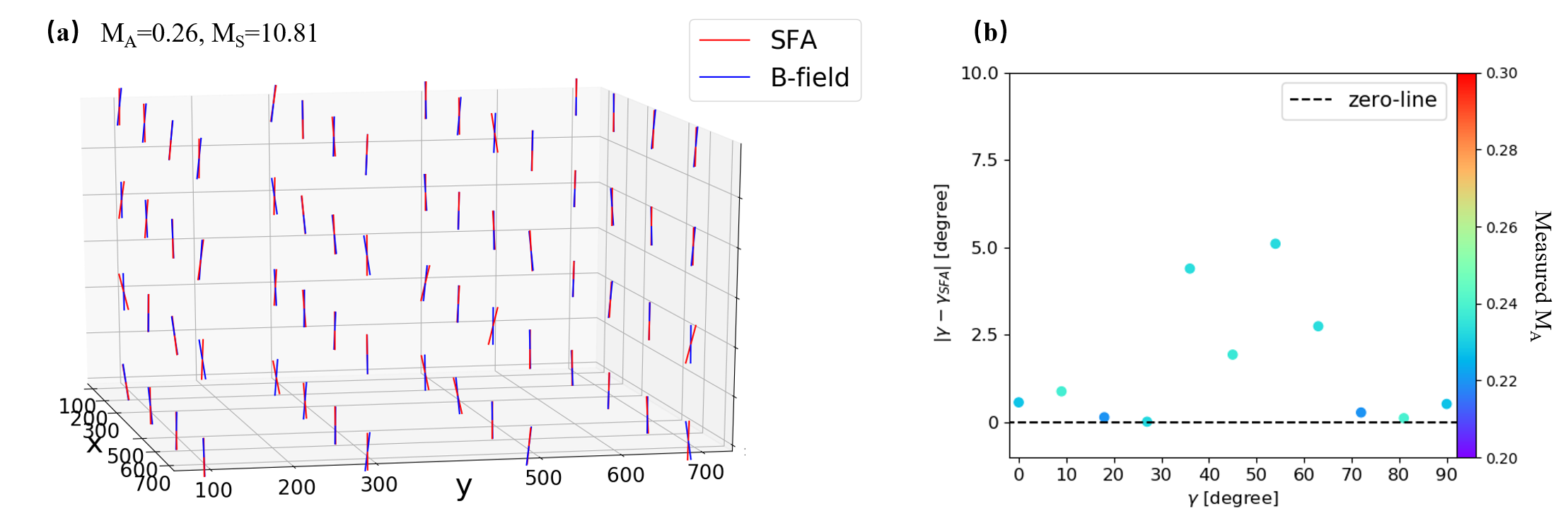}
\caption{\label{fig:gamma} \textbf{Panel a:}
An example of the 3D magnetic field traced by SFA (red segments) comparing with the real 3D magnetic fields (blue segments) in numerical simulation. Mean magnetic field is perpendicular to the LOS (x-axis), i.e., $\gamma=\pi/2$. \textbf{Panel b:} The relative angle of numerically measured mean $\gamma_{SFA}$ and real mean $\gamma$ (x-axis). Color of the point represents the numerically measured mean $\rm M_A$ for the simulation $\rm M_S=10.81$, $\rm M_A=0.26$. The dashed line represents the equality between the measured and real values.}
\end{figure*}

\section{Observational Applications}
\label{sec:obs}
\subsection{Application to young stars}
We give an example of using SFA to trace the 3D magnetic field. Here we use the simulation $\rm M_S\approx10.81$, $\rm M_A\approx0.26$, in which the mean magnetic field is inclined to the LOS with angle $\gamma=\pi/2$. We split the $792\times792\times 792$ velocity cube into 64 sub-cubes. Each sub-cube contains $198\times198\times198$ data points. Following the recipe presented in \S~\ref{subsec:sfa}, for each sub-cube, we vary the 3D direction of $\hat{r}$ in space and calculate its corresponding velocity structure-function $SF_2(v)$ along $\hat{r}$. We randomly select $1\times 10^5$ data points for the calculation. The direction of $\hat{r}$ corresponding to the minimum value of $SF_2(v)$ at separation $\approx0.1$ pc gives the measurement of 3D magnetic field direction. As shown in Fig.~\ref{fig:gamma}, the 3D magnetic fields measured from SFA agree with the real magnetic fields in the simulation. Both of them have a mean magnetic field approximately perpendicular to the LOS.

Besides, we rotate the simulation so that the mean magnetic field is inclined to the LOS with angle $\gamma\ne\pi/2$. Similarly, we also vary the direction of $\hat{r}$ in space and calculate its structure-function of velocity.  The direction corresponding to the minimum value of the structure-function (i.e., $v^2_{l,2}$) reveals the magnetic field orientation, while the direction with maximum structure-function (i.e., $v^2_{l,1}$) is perpendicular to the magnetic field. Then we derive $\rm M_A$ from:
\begin{equation}
    {\rm M_A}=(v_{l,1}^2/v_{l,2}^2)^{-3/4}
\end{equation}
Fig.~\ref{fig:gamma} presents the difference of numerically measured mean $\gamma_{SFA}$ from the SFA and real mean $\gamma$. The relative angle of $\gamma_{SFA}$ and $\gamma$ is smaller than $\approx5^\circ$, which represents an excellent alignment. The real $\rm M_A=0.26$ and we have the measured $\rm M_A\approx0.24$. The full 3D magnetic field information gets retrieved.

The SFA usually requires the knowledge of 3D positions, which is challenging to obtain in observation. Nevertheless, the flow motion can be traced by placing point-like objects into the turbulent flow \citep{2001Natur.409.1017L}. It means the calculation of the structure-function is applicable to point sources, which are similar to the points selected by the Monte Carlo method in \S~\ref{sec:results}. Therefore, one way to measure turbulent velocities is using velocities of point sources, as their motions are coupled to the background turbulent motions, or they inherit the turbulent motions from the parent turbulent gas in the case of young stars (see \citealt{Ha20}). The 6D phase-space information of stars can be obtained from the Gaia survey \citep{2016A&A...595A...1G,2018A&A...616A...1G}. By calculating the velocity fluctuations of stars, the SFA could reveal the 3D magnetic fields, including both LOS and POS components. 

One important question is how many point sources are required to give an accurate measurement. In Fig.~\ref{fig:thickness}, we vary the number of point sources used for a single bin in SFA's calculation. We consider the bin centred at 0.1 pc (spans from 0.09 pc to 0.11 pc) as above. We see the resulting ratio of velocity fluctuations $v_{l,1}^2/v_{l,2}^2$ gets statistically stable when the sample size is larger than one thousand. The required number of point source also depends on the property of the turbulent system. In observation, usually all point sources fall into the inertial range. It is likely the turbulent properties can be reflected by fewer point sources. In Fig.~\ref{fig:thickness}, we also investigate the effect of noise by introducing the Gaussian noise to the velocity field. Its amplitude varies from 0\% to 30\% of mean velocity. We selected one million points still. We see the ratio $v_{l,1}^2/v_{l,2}^2$ gets decreasing in the presence of noise. Comparing with the theoretical value $\approx1.7$, the SFA still works when the noise level is less than 10\%. Note here the amplitude of noise depends on the mean velocity, which is also the injection velocity at injection scale. The scale-depend velocity fluctuation (see Fig.~\ref{fig:SF_sub02}) at small scale is more sensitive to noise. Therefore, in practice, one should use the velocity fluctuations at large scales to suppress the noise.

\subsection{Application to spectroscopic data}
\begin{figure}
\centering
\includegraphics[width=1.0\linewidth,height=1.45\linewidth]{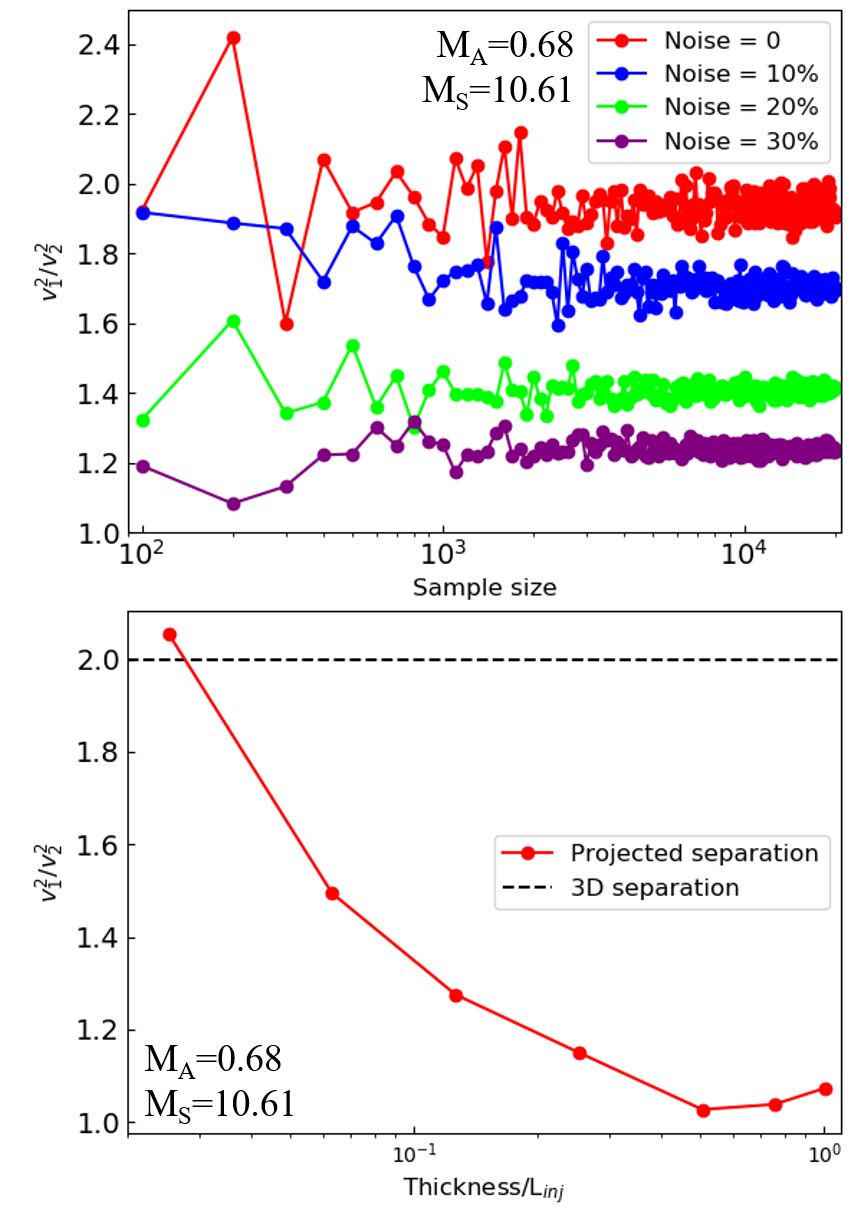}
\caption{\label{fig:thickness}\textbf{Top:} The correlation of measured $v_{l,1}^2/v_{l,2}^2$ and the sample size. \textbf{Bottom:} The correlation of measured $v_{l,1}^2/v_{l,2}^2$ and the thickness of cloud. $L_{inj}$ is the injection scale.}
\end{figure}

The SFA can also apply to dense molecular cores, which has volume density more than an order of magnitude higher than its surroundings \citep{2018ApJ...864..116Q}. The emission lines, i.e., the PPV cubes, of dense core provide the information of the POS projected separation and LOS velocity.  For instance, we can define the thickness of a cloud as its length scale along the LOS. In the case of its thickness being far smaller than its length scale on the POS, the cloud is thin. Otherwise, it is a thick cloud. In the limit of very thin cloud, the projected separation calculated approximately equivalents to the real separation between two dense cores in 3D. For instance, \citet{2015ApJ...811...71Q} found  Taurus is a nearly face-on thin molecular cloud and \citet{2018ApJ...864..116Q}
successfully obtained the structure function of velocities using Taurus' dense cores. In Fig.~\ref{fig:thickness}, we test the similar idea for the SFA. We vary the thickness of the cloud from 0.1 pc to 5.0 pc and use only the projected separation and LOS component of velocity. $1\times10^5$ points are selected by Monte Carlo method for the analysis. We find the measured velocity fluctuation ratio decreases for a thick cloud. When the cloud is thinner than 3\% of the injection scale, the measured value is close to the theoretical value. Therefore, as we expected, the SFA is applicable to dense molecular cores when the cloud is thin. Note searching for dense cores usually requires optically thin emission lines, which trace high-density regions \citep{2018ApJ...864..116Q}.

\section{Discussion}
\label{sec.dis}
\subsection{Tracing 3D magnetic fields and scale-dependent anisotropy}
\label{subsec:gaia}
The anisotropy of turbulence is widely observed in the ISM, including transparent diffuse gas and dense molecular gas \citep{2008ApJ...680..420H,2016A&A...595A..37K}. \citet{1983A&A...122..282C} and \citet{2016A&A...595A..37K} found that \ion{H}{1} emission's power spectra exhibit a typical slope value around $-3$, which is independent of position angle. In particular, \citet{2016A&A...595A..37K} observed that the spectral power in the direction perpendicular to the magnetic field inferred from Planck polarization gets maximum value. This observational evidence confirms our theoretical expectation and numerical results.

In this work, we propose a novel approach to probe the magnetic fields and estimate magnetization using the second-order structure function of turbulent velocity, i.e. the Structure-Function Analysis (SFA). The SFA is observationally motivated by turbulence measurements with point sources. For example, \citet{Ha20} found that the velocities of young stars formed in a molecular cloud show signatures of turbulence  of the parent cloud. The velocity field sampled by stars is expected to be anisotropic in the presence of magnetic fields and thus contains the magnetic field information. As we discussed in \S~\ref{sec:theory}, the ratio of the velocity fluctuations measured in the parallel and perpendicular directions (with respect to the mean magnetic field) has a power-law dependence on Alfv\'{e}n Mach number $\rm M_A$. In this paper, we discuss the application of the SFA to the turbulent velocities measured with point sources, e.g., dense molecular cores \citep{2018ApJ...864..116Q}, dense filaments \citep{2020ApJ...889L...1L}, young stars \citep{Ha20}. As the Gaia survey \citep{2016A&A...595A...1G,2018A&A...616A...1G} provides 3D positions and 3D velocities of young stars, the SFA can directly employ this information to reveal the 3D magnetic fields. 

Note that the global reference frame discussed above (see \S~\ref{sec:theory}) is defined in terms of a given size of the sample turbulent volume. Multi-scale magnetic fields can be measured by varying length scales. The SFA therefore provides the possibility to study the predicted scale-dependent anisotropy in the local reference frame. The full 6D phase-space information of stars from the Gaia survey, together with the SFA, can simultaneously give the local 3D magnetic fields and 3D turbulent velocities. The magnetic field information defines the local reference frame. By employing the approach proposed by \citet{2000ApJ...539..273C}, the velocity fluctuations calculated in the local reference frame will exhibit a scale-dependent anisotropy. In addition, \citet{2003MNRAS.345..325C} decompose the turbulence into fast, slow, and Alfv\'{e}nic modes in the local reference frame. The similar idea of defining the local reference frame through the Gaia survey and SFA can also be implemented in the decomposition method. 

In addition, turbulence and magnetic fields play important roles in the formation of density filaments \citep{2018A&A...614A.100T,2019A&A...632A..68T,2019ApJ...878..157X} and the origin of core-mass function \citep{2012MNRAS.423.2037H,2019FrASS...6....7K}. Our study provides a unique way to trace 3D magnetic fields through turbulence's anisotropy. It may give new insight into explaining the balance between turbulence, magnetic fields, and gravity in these processes.

\subsection{Application with 2D velocity information}
The application to probe the 3D magnetic fields associated with stars is straightforward. The Gaia survey provides 3D position and 3D velocity information for calculating the structure function. However, obtaining the velocity information of either multi-phase atomic gas or cold molecular gas is non-trivial in observations. For sub-Alfv\'{e}nic turbulence, the velocity fluctuation is also preserved with the averaging along the LOS \citep{2005ApJ...631..320E,2017MNRAS.464.3617K}. Thus, one of the most common ways to extract velocity information is using the Position-Position-Velocity (PPV) cube to obtain velocity centroid map. A 2D velocity centroid map C(x,y) can be produced by integrating the PPV cube along the LOS:
\begin{equation}
    C(x,y)=\frac{\int v\cdot\rho(x,y,v)dv}{\int\rho(x,y,v)dv}
\end{equation}
where $v$ is the LOS component velocity at the position (x,y), including both the turbulent velocity and the residual velocity due to thermal motions. $\rho(x,y,v)$ is the gas density or radiation temperature. The integration eliminates the effect of temperature so that C(x,y) is only regulated by gas density and turbulent velocity \citep{2005ApJ...631..320E,2017MNRAS.464.3617K,H2}. The statistics of velocity centroid has been earlier developed to reveal the anisotropy \citep{2002ASPC..276..182L,2005ApJ...631..320E,2011ApJ...740..117E,2014ApJ...790..130B,2017MNRAS.464.3617K}. Therefore, the SFA is also applicable to the 2D velocity information provided by velocity centroid \citep{XH21}. This provides a possibility to test the Gaia related results (see \S~\ref{subsec:gaia}) with spectroscopic data.

The second way to extract velocity fluctuations comes from the velocity caustic effect in PPV cube \citep{LP00}. This effect reveals that turbulent velocities along the LOS can significantly distort the density structure. The distorted density structure is dominated by velocity fluctuations instead of density fluctuations. \citet{LP00} quantified the significance of velocity caustics in PPV cubes in terms of the density spectral index. When the density power spectrum is steep, i.e., the slope is shallower than -3, the velocity fluctuation dominates the PPV cube's emissivity spectrum. The importance of this velocity fluctuation is correlated to the width of the velocity channel $\Delta v$. When the channel is thin, i.e., $\Delta v<\sqrt{\delta(v^2)}$ where $\sqrt{\delta(v^2)}$ is the velocity dispersion, the velocity fluctuations are most prominent in corresponding velocity channels. When the channel is thick, i.e., $\Delta v>\sqrt{\delta(v^2)}$, density fluctuations are dominated the velocity channels \citep{LP00}. Several authors \citep{2019ApJ...874..171C,2020arXiv200301454K,2020ApJ...899...15M} argue that Cold Neutral Medium (CNM) is responsible for shallower spectral index observed in the thin velocity channel maps of \ion{H}{1} emission rather than the velocity caustics effect. We stress that the velocity caustics is a natural effect of nonlinear spectroscopic mapping from real PPP space to PPV space, which also inevitably presents in CNM. This issue will be discussed in detail elsewhere.

In multi-phase ISM, the velocity information provided by either the velocity centroid map or thin velocity channel is partially contaminated by density's contribution. For subsonic or trans-sonic warm ionized medium (WIM, \citealt{1987ASSL..134...87K,2008ApJ...686..363H}) and
warm neutral medium (WNM, \citealt{2010ApJ...714.1398C}), the compression of turbulence is insignificant so that density statistics passively follow velocity statistics \citep{2002ApJ...564..291C,2003MNRAS.345..325C,2019ApJ...878..157X}. Consequently, the velocity centroid map and thin velocity channel can directly reveal velocity fluctuations in WIM and WNM. When the turbulent compression is significant, i.e., for supersonic CNM and molecular clouds \citep{1974ARA&A..12..279Z,1981MNRAS.194..809L,2010ApJ...714.1398C}, the situation is more complicated. Density fluctuations here are subjected to not only the Alfv\'{e}nic mixing but also the shock compression. The latter results in a shallow density spectrum \citep{2005ApJ...624L..93B,2007ApJ...658..423K}. To eliminate the shock effect, one could select the velocity channel width as thin as possible, as the significance of velocity fluctuations is negatively related to the channel width.

\subsection{Comparison with other works}
The anisotropic properties of turbulence have been widely used to study magnetic fields. The first attempt was achieved by the Correlation Function Analysis (CFA) of velocity \citep{2002ASPC..276..182L,2011ApJ...740..117E}. CFA was further extended to determine magnetization \citep{2011ApJ...740..117E,2015ApJ...814...77E} and the contribution of the fast, slow, and Alfv\'{e}n modes in observed turbulence \citep{2016MNRAS.461.1227K,2017MNRAS.464.3617K}.  
\citet{2008ApJ...680..420H} provides a different way to trace magnetic fields using the Principal Component Analysis of Anisotropies (PCAA). The Velocity Gradients Technique (VGT; see \citealt{2017ApJ...835...41G,YL17a,LY18a,PCA,survey}) also employs the velocity anisotropy calculated for a given separation between the points. The largest velocity gradient is perpendicular to the eddy's orientation so that the velocity gradient is perpendicular to the magnetic fields. Similar to the present technique, in \citet{LY18a} the difference between the maximum and minimum velocity gradient was used to obtain $\rm M_A$. The advantage of the SFA is that using the 3D information on turbulent velocities and distances, one can obtain the actual 3D magnetization. 

We note that the VGT approach can also be implemented with intensity gradients, i.e. without the velocity information \citep{YL17b,IGs,cluster}. In this case, one can obtain $\rm M_A$ employing the \citet{2018ApJ...865...46L} approach and relating the dispersion of intensity gradients with $\rm M_A$ to estimate the magnetization. In general, these approaches detect the eddy's contour. Since the eddy is elongated along with the local magnetic field, the semi-major axis reveals the magnetic field direction. These techniques mentioned above usually employ spectroscopic data to extract the turbulent velocity of plasma gas in scales from protostars to galaxy clusters. Their application for velocities of point sources is not yet explored. The SFA, however, instead of directly detecting the eddy's contour, focuses on the velocity of point sources. The amplitude of velocity fluctuation gets maximum when the measuring direction is perpendicular to the magnetic fields. Also, the anisotropy degree, which correlates with magnetization in a power-law behavior, can be used to estimate the magnetization.

An important advantage of SFA is that it requires much less valid data points. For instance, the selected one million points occupy only a small fraction $10^6/792^3\approx0.2\%$ of the entire cube. This method directly calculates the velocity fluctuation between each pair of points. At a given separation, the velocity fluctuation gets its minimum value in the direction parallel to the magnetic fields. It naturally raises a question: how many samples do we need to have a statistically stable velocity fluctuations at a fixed separation? The answer is presented in Fig.~\ref{fig:thickness}, in which we vary the number of points used to calculate the velocity fluctuation with the separation $\approx$ 0.1 pc. We found the velocity fluctuation becomes stable when the sample size is larger than approximately 1000 points. The idea of finding a stable velocity fluctuation at a fixed separation is similar to the sub-block averaging method implemented in VGT. The sub-block averaging method uses a statistically stable histogram of velocity gradients to find the direction of magnetic fields \citep{YL17a}.

The CFA method detects a compact iso-contour of velocity fluctuations. The semi-major-axis of the iso-contour indicates the direction of the magnetic field. If the correlation-function is calculated directly between each pair of points. The required number of point sources to find a single velocity iso-contour for CFA and a single-separation iso-contour for SFA is similar. However, since the value of velocity iso-contour cannot directly be fixed in observation, one has to try a number of point sources at different separations. This results in multiple iso-contours of velocity fluctuations and significantly increases the required data number. \cite{2018ApJ...865...54Y} use the open-boundary Fast Fourier Transform (FFT) to improve the calculation of CFA. The FFT decreases the required computational time. However, because the FFT is also sensitive to the number of data points, the FFT-based CFA requires even more data points than directly computing the correlation function. As shown in \citet{2018ApJ...865...54Y}, this improved CFA requires at least 60\% valid data points of the entire system to trace the magnetic fields. Comparing with the CFA method, the iso-contour of separations required by SFA are directly available in observations. One can easily fix the two-point separation value to obtain a single iso-contour of separations. Therefore, SFA requires fewer data points in practice and gives advantageous in probing the magnetic fields associated with point sources and dealing with low-resolution or sparse-pixel spectroscopic data. 


\section{Summary} 
\label{sec:con}
Anisotropy is an intrinsic property of sub-Alfv\'{e}nic MHD turbulence. A number of studies have employed anisotropy to study the magnetic field in the ISM. In this work, based on also the turbulence anisotropy, we propose the Structure-Function Analysis (SFA) as a novel approach to probe the magnetic fields and estimate the magnetization. This new approach is complementary to the other ways of tracing the magnetic field. Our main results are summarized as follows.
\begin{enumerate}
    \item Based on the analytical model of MHD turbulence and numerical experiments, we showed that in both the local and global reference frames, the velocity fluctuations are the most significant in the direction perpendicular to the magnetic field.
    \item We find that the ratio of velocity fluctuations in perpendicular and parallel directions with respect to magnetic field direction has a power-law relation with $\rm M_A$, which is inversely proportional to magnetic field strength. 
    \item We show that the turbulent velocities traced by point sources can be used to study the 3D magnetic fields. With turbulent velocities measured by spectroscopic data, the SFA can probe the magnetic fields in multi-phase interstellar medium. 
    \item We discuss the possibility of tracing 3D magnetic fields and actual magnetization using the SFA and the Gaia survey.
\end{enumerate}

\section*{Acknowledgements}
 Y.H. acknowledges the support of the NASA TCAN 144AAG1967. S.X. acknowledges the support for this work provided by NASA through the NASA Hubble Fellowship grant \# HST-HF2-
51473.001-A awarded by the Space Telescope Science Institute, which is operated by the Association of Universities for Research in Astronomy, Incorporated, under NASA
contract NAS5- 26555. A.L. acknowledges the support of the NSF grant AST 1715754 and NASA ATP AAH7546.
\software{Julia \citep{2012arXiv1209.5145B}, ZEUS-MP/3D code \citep{2006ApJS..165..188H}}




\end{document}